\documentclass[12pt]{article}

\usepackage[left=2.8cm,right=2.8cm,top=2.9cm,bottom=2.9cm]{geometry}

\usepackage[colorlinks=true,linkcolor=black,citecolor=blue,urlcolor=blue,filecolor=black]{hyperref}

\usepackage[nosort]{cite}
\usepackage{latexsym}
\usepackage{amsmath}
\usepackage{amsfonts}
\usepackage{ytableau}

 \csname
@addtoreset\endcsname{equation}{section}

\def\cL{{\cal L}}

\def\be{\begin{equation}}
\def\ee{\end{equation}}

\def\12{\frac{1}{2}}

\begin{document}

\begin{center}

%%%%%%%%%%%%%%%%%%%%%%%%%%%%%%%%%%%%%%%%%%%%%%%%%%%%%%%%%%%%%%%%%%%%
{\Large\bf Spin-2 twisted duality in (A)dS} 
%%%%%%%%%%%%%%%%%%%%%%%%%%%%%%%%%%%%%%%%%%%%%%%%%%%%%%%%%%%%%%%%%%%%

\vspace{1cm}

Nicolas~Boulanger${}^{\, a}$, Andrea~Campoleoni${}^{\,a, b,}$\footnote{Research Associate of the Fund for Scientific Research-FNRS (Belgium).}, Ignacio~Cortese${}^{\, c}$ and Lucas~Traina${}^{\, a,}$\footnote{Research Fellow of the Fund for Scientific Research-FNRS (Belgium).}

\vspace{.8cm}

{${}^a$\sl\footnotesize
Service de Physique de l'Univers, Champs et Gravitation,\\  \mbox{Universit\'e de 
Mons -- UMONS}, 20 place du Parc, 7000 Mons, Belgium}
\vspace{5pt}

{${}^b$\sl\footnotesize
Institut f\"ur Theoretische Physik, ETH Zurich,\\
Wolfgang-Pauli-Strasse 27, 8093 Z\"urich, Switzerland
}
\vspace{5pt}

{${}^c$\sl\footnotesize
Departamento de F{\'\i}sica de Altas Energ{\'\i}as, \mbox{Instituto de Ciencias Nucleares -- UNAM},\\
Circuito Exterior s/n, Cd.\ Universitaria, 04510 Ciudad de M\'exico, Mexico
}
\vspace{5pt}

{\tt\footnotesize \href{mailto:nicolas.boulanger@umons.ac.be}{nicolas.boulanger@umons.ac.be}, \href{mailto:andrea.campoleoni@umons.ac.be}{andrea.campoleoni@umons.ac.be},\\ \href{mailto:nachoc@nucleares.unam.mx}{nachoc@nucleares.unam.mx}, \href{mailto:lucas.traina@umons.ac.be}{lucas.traina@umons.ac.be}}

\end{center}

\vspace{1cm}

\noindent Starting from the dual Lagrangians recently 
obtained for (partially) massless spin-2 fields in the 
Stueckelberg formulation, we write the equations of motion 
for (partially) massless gravitons in (A)dS in the form of 
twisted-duality relations.
In both cases, the latter admit a smooth flat limit. 
In the massless case, this limit reproduces 
the gravitational twisted-duality relations 
previously known for Minkowski spacetime.
In the partially-massless case, our twisted-duality relations
preserve the number of degrees of freedom in the flat limit,
in the sense that they split into a decoupled pair of dualities 
for spin-1 and spin-2 fields.
Our results apply to spacetimes of any dimension greater than 
three. In four dimensions, the twisted-duality relations for 
partially massless fields that appeared in the literature are 
recovered by gauging away the Stueckelberg field.

%\tableofcontents

\newpage

%%%%%%%%%%%%%%%%%%%%%%%%%%%%%%%%%%%%%%%%%%%%%%%%%%%%%%%%%%%%%%%%
\section{Introduction and conventions}\label{sec:conventions}
%%%%%%%%%%%%%%%%%%%%%%%%%%%%%%%%%%%%%%%%%%%%%%%%%%%%%%%%%%%%%%%%

Electric-magnetic duality, the 
symmetry of vacuum Maxwell equations under the exchange of electric 
and magnetic fields that interchanges dynamical equations with Bianchi identities, 
has counterparts in other physical systems, including supersymmetric field theories, 
linearised gravity and free higher-spin gauge theories. 
In supersymmetric Yang-Mills theories, electric-magnetic duality 
--- see \cite{Olive:1999bx} and references therein --- acts as a 
strong/weak duality and, as such, has found applications in the study of 
non-perturbative phenomena like confinement (see e.g.\ \cite{Strassler}). 
In extended supergravity theories it is part of the U-duality symmetry and, 
since the pioneering work \cite{Cremmer:1979up}, it has been extensively studied.

If the dimension of spacetime is bigger than four,  electric-magnetic duality 
actually links different descriptions of the same physical system. 
For linearised gravity on a flat background in $n$ dimensions, for instance, 
it relates the Fierz-Pauli description in terms of the tensor $h_{ab}$ with 
a description in terms of an irreducible mixed-symmetry tensor 
$T_{a_1 \dots a_{n-3} | b}$, completely antisymmetric in its first $n-3$ indices. 
This link has however been established only at the linearised level: 
non-linear Einstein gravity cannot be reproduced in the dual mixed-symmetry 
picture by means of local interactions \cite{Bekaert:2002uh,Bekaert:2004dz}.
The problems encountered in the attempts to lift gravitational dualities from 
the linearised formulation in flat background to the interacting level 
suggest that the study of electric-magnetic duality in curved backgrounds 
may be particularly promising. 
Positive results about interaction vertices for mixed-symmetry fields are indeed 
available in this context \cite{Boulanger:2011se} 
and may indicate a way to extend the duality to the interacting theory. 

Recently \cite{Boulanger:2018shp}, manifestly covariant action 
principles in the Stueckelberg formulation were given for dual 
massless, partially 
massless and massive spin-2 fields in maximally symmetric spacetimes 
of arbitrary dimensions $n>3\,$, such that 
the degrees of freedom are preserved in the flat 
limit.
The action principles for the dual fields were also related 
to the standard ones for such field theories by building on the 
previous works
\cite{Curtright:1980yj,Fradkin:1984ai,West:2001as,
Zinoviev:2001dt,West:2002jj,Boulanger:2003vs,
Matveev:2004ac,Zinoviev:2005zj,Zinoviev:2005qp,Gonzalez:2008ar,
Khoudeir:2008bu,Basile:2015jjd}. See also \cite{Mignaco:2001pi} for references 
on earlier works.

In this Letter we focus on the massless and 
partially-massless cases and formulate the field equations 
derived from the actions of \cite{Boulanger:2018shp} 
as twisted-duality relations. 
In the massless case, our twisted-duality relation
---~see Eq.~\eqref{twisted_dual}~--- generalises to 
(A)dS backgrounds the twisted-duality relation 
written in \cite{Hull:2000zn,Hull:2001iu} for linearised Einstein 
gravity in flat spacetimes.
Our duality relation actually smoothly reproduces the latter
duality relation in the flat limit, thanks to the crucial 
role played by the Stueckelberg fields.

In the case of a partially-massless spin-2 field 
\cite{Deser:1983mm}, 
the twisted-duality relation that we obtain ---~see Eq.~\eqref{twisted_dual_pm1}~--- 
has a smooth flat limit that reproduces a couple of 
twisted-duality relations in flat background, 
one for a massless spin-2 field and the other for a 
massless spin-1 field, thereby correctly accounting for the 
degrees of freedom of a partially-massless spin-2 field. 
Moreover, keeping the cosmological constant non-zero
and setting the dimension of spacetime to $n=4\,$, 
our twisted-duality relation reproduces the one 
given in \cite{Hinterbichler:2014xga}, 
upon eliminating the Stueckelberg field. 

Twisted-duality relations are interesting for many reasons. 
In particular they 
relate, for a pair of dual theories, the Bianchi identities 
of one system to the field equations of the dual one, 
and vice versa. In the present work, we show that the 
field equations of two dual theories are formulated as
a twisted-duality equation, although we note that 
the latter is not obtained from a variational principle
that is manifestly spacetime covariant. 
Forgoing the latter requirement, for
linearised Einstein theory around flat spacetime
the authors of \cite{Bunster:2013oaa} gave an 
action principle that yields the twisted self-duality 
conditions as equations of motion, keeping the graviton and its 
dual on equal footing.
Finally, let us mention that, for the fully nonlinear 
Einstein-Hilbert theory, 
an action principle was given in \cite{Boulanger:2008nd} 
where both the graviton and its dual appear inside the action, 
albeit not on an equal footing and together 
with extra auxiliary fields.
For recent interesting works where twisted (self) duality 
relations play a central role and for more references, 
see  
\cite{Henneaux:2016opm,Henneaux:2017xsb,Henneaux:2018rub,Lekeu:2018vhq}.

As for our conventions, we work on constant-curvature spacetimes 
with either negative or positive cosmological constant 
$\Lambda\,$. 
We denote the number of spacetime dimensions by $n$ 
and define the quantity
$\lambda^2 = \tfrac{-\,2\,\sigma\,\Lambda}{(n-1)(n-2)}\,$, 
$\sigma = \pm 1\,$, that is always positive. 
When the background is AdS$_n$ one has $\sigma = 1\,$, 
while $\sigma = -1$ for dS$_n\,$.
The commutator of covariant derivatives gives
$[ \nabla_{\!a} , \nabla_{\!b} ] V_c = -\, 
\sigma \lambda^2 \left( g_{ac} V_b - g_{bc} V_a \right)\,$,
where $g_{ab}$ is the background (A)dS$_n$ metric. 
The symbols 
$\epsilon_{a_1\cdots a_n}$ and $\epsilon^{a_1\cdots a_n}$ 
denote the totally antisymmetric tensors obtained from 
the corresponding densities upon multiplication and division 
by $\sqrt{-g}\,$.

%%%%%%%%%%%%%%%%%%%%%%%%%%%%%%%%%%%%%%%%%%%%%%
\section{Massless spin-2 twisted duality}
%%%%%%%%%%%%%%%%%%%%%%%%%%%%%%%%%%%%%%%%%%%%%%

%***********************************************************
\subsection{Fierz-Pauli formulation}
%***********************************************************

In the Fierz-Pauli formulation for a massless spin-2 field
around a maximally-symmetric spacetime of dimension $n\,$, 
the Lagrangian
(where we omit the factor $\sqrt{-{g}}$ for the sake of 
conciseness) is given by
\begin{equation}\label{LFP}
\begin{split}
{\cal L}^{FP} =& 
-\tfrac{1}{2}\,\nabla_{\!a} h_{bc}\nabla^a h^{bc}  
+ \nabla_{\!a} h_{bc}\nabla^c h^{ba} 
+ \tfrac{1}{2}\,\nabla_{\!a} h \nabla^a h 
- \nabla_{\!a} h\nabla_b h^{ab} \\ 
& -\tfrac{(n-1)\sigma\lambda^2}{2} 
\left(2\,h_{ab}h^{ab}-h^2\right)\,.
\end{split}
\end{equation}
It is invariant, up to a total derivative, under the gauge
transformations\footnote{Indices enclosed between (square) 
round brackets are (anti)symmetrised, and dividing by 
the number of terms involved is understood (strength-one 
convention). Moreover, we will use a vertical bar to separate 
groups of antisymmetrised indices, see e.g.~Eq.~\eqref{FPcurvature}.}
\begin{align}
    \delta h_{ab} = 2\,\nabla_{\!(a}\xi_{b)}\;.
\end{align}
The primary gauge-invariant quantity for the Fierz-Pauli
theory is given by 
\begin{equation} \label{FPcurvature}
\begin{split}
K^{ab|mn} =\;&
-\tfrac{1}{2}\,\left(
\nabla^a \nabla^{[m} h^{n]b} - \nabla ^b 
\nabla^{[m} h^{n]a} + \nabla ^m \nabla^{[a} h^{b]n} 
- \nabla ^n \nabla^{[a} h^{b]m}\right) \\
& \qquad + \sigma\lambda^2 \left({g}^{a[m} h^{n]b} 
- {g}^{b[m} h^{n]a} \right)\,.
\end{split}
\end{equation}
It possesses the same symmetries as the components of the Riemann 
tensor,
\begin{align}
 K_{[ab|c]d} &\equiv 0\;,
 \label{b1}
\end{align}
and obeys the differential Bianchi identity
\begin{align}
    \nabla^{[a}K^{b c]|mn}& \equiv  0\;.
    \label{bianchidiff}
\end{align}
The field equations derived from the Lagrangian ${\cal L}^{FP}$ 
imply the tracelessness of the curvature: 
\begin{align}
    K_{mn}:= g^{a b}\,K_{m a|n b}
    \approx 0\;,
    \label{einstein1}
\end{align}
where weak equalities are used throughout this paper
to indicate equalities that hold on the surface of the solutions
to the equations of motion.
More precisely, defining \mbox{$K= g^{ab}K_{ab}\,$}, 
the left-hand side of the field equations read
\begin{align}
    \frac{\delta {\cal L}^{FP}}{\delta h^{ab}}\; \equiv 
    -2 \left(K_{ab} - \tfrac{1}{2}\,g_{ab}K \right)\,.
\end{align}
By virtue of the differential Bianchi identity for the curvature,
one also finds that, on-shell, the curvature has vanishing divergence:
\begin{align}
    \nabla^m K_{m n|ab} \approx 0\;.
    \label{einstein2}
\end{align}

To summarise, the important equations in this section 
are \eqref{b1}, \eqref{bianchidiff} and \eqref{einstein1}.
The latter relation was derived from the Lagrangian 
${\cal L}^{FP}\,$. For the purpose of deriving a twisted-duality
relation, we can actually forget the origin of \eqref{einstein1}
and focus on the three equations \eqref{b1}, \eqref{bianchidiff} 
and \eqref{einstein1}.

%---------------------------------------------------
\subsection{Dual formulation}
\label{sec:stueckelberg:massless}
%---------------------------------------------------

We start from the dual formulation of the massless 
spin-2 theory as given by the Lagrangian 
${\cal L}_0({\hat Y},W)$ in Eq.~(20) of 
\cite{Boulanger:2018shp}: 
\be
\label{purespin2masslessandmassive}
\begin{split}
\cL_{0}({\hat Y},W) & = 
\tfrac{1}{\lambda^2} \Big[ \tfrac{1}{2}\,
\nabla_{\!c} W^{abc|d} \nabla^e W_{dbe|a} 
+ \lambda\,\hat{Y}^{ab|c} \nabla^e W_{cbe|a} \\
& \qquad + \tfrac{\sigma}{2(n-2)}\, \nabla_{\!b} 
\hat{Y}^{ab|c} \nabla^d \hat{Y}_{cd|a}  + \tfrac{\lambda^2}{2}\,
\hat{Y}^{ab|c} \hat{Y}_{ac|b} \Big] \; . \\
\end{split}
\ee
This Lagrangian describes the propagation of the same degrees 
of freedom as the Fierz-Pauli one in \eqref{LFP}. It has been built in \cite{Boulanger:2018shp} in two steps: a Lagrangian depending only on the field $\hat{Y}_{ab|c}$ (that is a traceless combination of the components of the spin connection) is obtained by eliminating the vielbein from the first-order formulation of linearised gravity in (A)dS. The full Lagrangian \eqref{purespin2masslessandmassive} then results from the Stueckelberg shift
\be \label{shiftY}
    \hat{Y}^{bc|}{}_a \, \to \, \hat{Y}^{bc|}{}_a 
    + \tfrac{1}{\lambda}\,\nabla_{\!d} W^{bcd|}{}_a\;.
\ee
The Lagrangian (2.9) possesses as many differential 
gauge symmetries as the Lagrangian obtained in \cite{Boulanger:2003vs} 
that describes dual (linearised) gravity in Minkowski background.

We now define the following quantities 
\begin{align}
    R_{ab|}{}^{cd} &:= 
    2\, \nabla_{\![a|}\! \left( \nabla_{\!e}W^{cde}{}_{|b]}
    + \lambda \, \hat{Y}^{cd}{}_{|b]} \right)\,,
    \\[5pt]
    K_{ab|}{}^d & := 
    2\,\nabla_{\![a|}\nabla_{c}\hat{Y}^{cd}{}_{|b]} 
    + 2\sigma(n-2) \lambda \left( \nabla_{\!c}W^{cd}{}_{[a|b]} + 
    \lambda \,\hat{Y}^{d}{}_{[a|b]} \right)\;,
\end{align}
together with their various non-vanishing traces
\begin{align}
    R_{a|}{}^{c} &= R_{ab|}{}^{cb}\;,\qquad 
    \qquad K_a = K_{ab|}{}^b\;.
\end{align}
Further introducing the traceless tensor $V_{ab|}{}^{cd}$ 
encoding the traceless projection of $R_{ab|}{}^{cd}\,$,
\begin{equation}
    V_{ab|}{}^{cd} = R_{ab|}{}^{cd} - 
    \tfrac{4}{n-2}\, \delta^{[c}{}_{[a}R_{b]|}{}^{d]}\;,
\end{equation}
we find that $V_{ab|}{}^{cd}$ is invariant
under the following gauge transformations:
\begin{align}
\delta \hat{Y}^{bc|}{}_{a} &= \nabla_{\!d} \zeta^{bcd|}{}_{a} 
+ \nabla_{\!a} \Lambda^{bc} + 
\tfrac{2}{n-1}\, \delta_a{}^{[b} \nabla_{\!d} \Lambda^{c]d} 
+ (n-3)\sigma\lambda\, \chi^{bc}{}_{a} \;,
\label{gaugevariatforYandWandhA}
\\[10pt]
\delta W^{bcd|}{}_{a} &= \nabla_{\!e} {\upsilon}^{bcde|}{}_{a}
+ \nabla_{\!a} \chi^{bcd} - \tfrac{3}{n-2}\, \delta_a{}^{[b} 
\nabla_{\!e} \chi^{cd]e} 
-\lambda\, \zeta^{bcd|}{}_{a} \;. \label{varW_gen}
\end{align}
Finally, the traceless tensor 
\begin{align}
    X_{ab|}{}^c &:= K_{ab|}{}^c + \tfrac{2}{n-1}\,\delta^c{}_{[a}K_{b]}
\end{align}
is also found to be gauge invariant.

As in \cite{Boulanger:2018shp}, one can also express the fields  
$W$ and $\hat{Y}$ in terms of their Hodge duals, that we denote 
by $C$ and $T$:\footnote{We substitute groups of antisymmetrised 
indices with a label denoting the total number of indices, 
e.g., $\epsilon_{a_1 \cdots a_n} \equiv \epsilon_{a[n]}\,$. 
Moreover, repeated indices denote an antisymmetrisation, e.g., 
$A_a B_a \equiv A_{[a_1} B_{a_2]}$.}
\begin{align} \label{HodgeDual}
 W^{abc|}{}_d &= -\tfrac{1}{(n-3)!}\,\epsilon^{e[n-3]abc}
 C_{e[n-3]|d}\;,
 &\hat{Y}^{ab|}{}_d = -\tfrac{1}{(n-2)!}\,\epsilon^{e[n-2]ab}
 T_{e[n-2]|d}\;.
\end{align}
The corresponding curvatures are obtained from the previous gauge-invariant tensors $V$ and $X$ as follows:
\begin{align}
K^C_{a[n-2]|bc} &= \tfrac{1}{2!}\,
\epsilon_{a[n-2]de}\,V_{bc|}{}^{de}\;,
&K^T_{a[n-1]|bc} = 
\epsilon_{a[n-1]d}\,X_{bc|}{}^d\;.
\end{align}
In components, the curvature tensors read
\begin{align}
 K^C_{a[n-2]|}{}^{bc} =\;& 2(n-2)(-1)^{n-1}\nabla^{[b}\nabla_{\!a}
 C_{a[n-3]|}{}^{c]} + 2\lambda\, \nabla^{[b}T_{a[n-2]|}{}^{c]}
 +\ldots\;,\\
 \nonumber \\
 K^T_{a[n-1]|}{}^{bc} =\;& 2(n-1)(-1)^{n}\nabla^{[b}\nabla_{\!a}
 T_{a[n-2]|}{}^{c]} - 2\sigma\lambda (n-1)(n-2)^2\,\delta^{[b}{}_a
 \nabla_{\!a} C_{a[n-3]|}{}^{c]}
 \nonumber \\
 & \quad + 2\sigma\lambda^2(n-1)(n-2)\,
 T_{a[n-2]|}{}^{[b}\delta_a{}^{c]} + \ldots\;,
\end{align}
where the ellipses denote terms that are necessary to ensure  
$GL(n)$-irreducibility of the curvatures 
$K^C{}_{a[n-2]|bc}$ and $K^T{}_{a[n-1]|bc}$ on the 
two-column Young tableaux of types $[n-2,2]$ 
 and $[n-1,2]\,$, 
respectively. Pictorially, they are represented by
\begin{equation*}
\ytableausetup
 {mathmode, boxsize=2}
\ytableausetup
 {mathmode, boxsize=2em}
\begin{ytableau}
 \scriptstyle a_1 & \scriptstyle b \\
 \scriptstyle a_2 & \scriptstyle c \\
\vdots\\
     \scriptstyle a_{n-2}
  \end{ytableau}
  ~~~~~\text{and}~~~~~
  \ytableausetup
 {mathmode, boxsize=2}
\ytableausetup
 {mathmode, boxsize=2em}
\begin{ytableau}
 \scriptstyle a_1 & \scriptstyle b \\
 \scriptstyle a_2 & \scriptstyle c \\
\vdots\\
     \scriptstyle a_{n-2} \\
     \scriptstyle a_{n-1}
  \end{ytableau}\quad .
  \end{equation*}
Indeed, tracelessness of $V_{bc|}{}^{de}$
and $X_{ab|}{}^c$ implies that the Hodge dual tensors 
$K^C_{a[n-2]|bc}$ and $K^T_{a[n-1]|bc}$ obey the following
algebraic Bianchi identities:
\begin{align}
    K^C_{a[n-2]|ac} &\equiv 0\;,\qquad 
    & K^T_{a[n-1]|ac} \equiv 0\;. \label{BI}
\end{align}
The two curvatures are linked via the following differential 
Bianchi identities:
\begin{align}
     \nabla_{\!a} K^C_{a[n-2]|}{}^{bc} \,&\equiv \, \lambda\,\tfrac{(-1)^n(n-3)}{(n-1)(n-2)}\,
     K^T_{a[n-1]|}{}^{bc}\;,
\label{BianchiII1} \\[10pt]
    \nabla^{[b} K^C{}_{a[n-2]|}{}^{cd]} \,& \equiv\,
    \lambda\,\tfrac{1}{n-2}\,K^T{}_{a[n-2]}{}^{[b|cd]}\;. 
\label{BianchiII2}    
\end{align}
These are equivalent to the following two identities:
\begin{align}
\nabla_{\!d} V_{ab|}{}^{cd} ~&\equiv~ -\lambda \,\tfrac{n-3}{n-2} \; X_{ab|}{}^{c}\;,
%\label{BianchiII1bis}
&\nabla_{\![b}V_{cd]|}{}^{ij} ~\equiv~ \lambda\,\tfrac{2}{n-2} \,
\delta_{[b}{}^{[i} X_{cd]|}{}^{j]}\;.
\label{BianchiII2bis}
\end{align}

The equations of motion for the dual 
gauge fields $C_{a[n-3]|b}$ and $T_{a[n-2]|b}\,$
derived from the Lagrangian 
${\cal L}_0(C,T)\,$ ---~obtained by substituting 
\eqref{HodgeDual} in \eqref{purespin2masslessandmassive} 
and given in Eq.~(23) of \cite{Boulanger:2018shp}~---  
can be written in terms of the traces of the gauge-invariant 
curvatures $K^C_{a[n-2]|b[2]}$ and $K^T_{a[n-1]|b[2]}\,$.
Explicitly, one has
\begin{align}
    \frac{\delta {\cal L}_0}{\delta C_{a[n-3]|b}} &\equiv 
    \tfrac{1}{\lambda^2(n-3)!}\,\Big(K_C{}^{a[n-3]c|}{}_{c}{}^b
    +\tfrac{n-3}{2}\, K_C{}^{a[n-4]cd|}{}_{cd}\,g^{ab}\Big)
    \approx 0\;,
    \label{EOM1}
    \\[5pt]
    \frac{\delta {\cal L}_0}{\delta T_{a[n-2]|b}} &\equiv 
    \tfrac{\sigma}{\lambda^2(n-2)^2(n-3)!}\,
    \Big( K_T{}^{a[n-2]c|}{}_{c}{}^b
    +\tfrac{n-2}{2}\, 
    K_T{}^{a[n-3]cd|}{}_{cd}\,g^{ab}\Big)\approx 0\;.
    \label{EOM2}
\end{align}
The field equations  \eqref{EOM1} and \eqref{EOM2} 
can easily be obtained by starting 
from the field equations of the Lagrangian 
${\cal L}_0(W,\hat{Y})$ and then expressing the fields 
$W^{abc|}{}_d$ and $\hat{Y}^{ab|}{}_c$ in terms of their Hodge
duals $C_{a[n-3]|b}$ and $T_{a[n-2]|b}\,$, respectively.
More in details, the left-hand sides of the 
field equations derived from ${\cal L}_0(W,\hat{Y})$ read
\begin{align}
    \frac{\delta {\cal L}_0(W,\hat{Y})}{\delta W^{abc|}{}_d}\, 
    &=
    \tfrac{1}{2\lambda^2}\,V_{[ab|c]}{}^d\;,
    &\frac{\delta {\cal L}_0(W,\hat{Y})}{\delta 
    \hat{Y}^{ab|}{}_d}\, 
    = - \tfrac{\sigma}{2(n-2)\lambda^2}\,X_{ab|}{}^d\;,
    \label{EOMforY}
\end{align}
and the gauge invariant tensors $X$ and $V$ can be expressed as 
\begin{align}
    X_{ab|}{}^d &= -\tfrac{1}{(n-1)!}\,\epsilon^{c[n-1]d}\,
    K^T{}_{c[n-1]|ab}\;,
    &V_{ab|}{}^{cd} = -\tfrac{1}{(n-2)!}\,\epsilon^{e[n-2]cd}\,
    K^C{}_{e[n-2]|ab}\;.
\end{align}

The field equations \eqref{EOM1} and \eqref{EOM2} imply
the tracelessness of the curvatures:
\begin{align}
    K_C{}^{a[n-3]c|}{}_{c}{}^b &\approx 0\;,
    &K_T{}^{a[n-2]c|}{}_{c}{}^b \approx 0\;.
    \label{EinsteinI}
\end{align}
In fact, from a result in representation theory of the 
orthogonal group --- see the theorem on p.~394 
of \cite{Hamermesh} ---, the second equation above implies that
\begin{align}
    K^T{}_{a[n-1]|bc} \approx 0\;.
    \label{ZeroKT}
\end{align}
The curvature for the field $T$ thus vanishes on shell, 
consistently with the observation that this field does not 
propagate any degrees of freedom in the flat limit 
\cite{Basile:2015jjd,Boulanger:2018shp}.

Upon using the first and second differential Bianchi identities 
\eqref{BianchiII1} and \eqref{BianchiII2}, we also find
the following two relations that are true on shell:
\begin{align}
     \nabla_{\!a} K^C_{a[n-2]|}{}^{bc} &\approx 0\;, 
     &\nabla^{[b} K^C_{a[n-2]|}{}^{cd]} \approx 0\;.
     \label{BianchiII}
\end{align}
These equations, together with \eqref{EinsteinI}, imply
that the divergences of the curvature $K^C$ vanish on shell: 
\begin{align}
     \nabla^a K^C_{ab[n-3]|cd} &\approx 0\;, 
     &\nabla^{b} K^C_{a[n-2]|bc} \approx 0\;.
     \label{EinsteinII}
\end{align}

To summarise, the important equations of this section 
are the equations of motion \eqref{EinsteinI} and the Bianchi 
identities \eqref{BI}, \eqref{BianchiII1} and \eqref{BianchiII2}. 
In the following section we will relate them to the field 
equations and the Bianchi identities of the Fierz-Pauli 
formulation via a twisted-duality relation. 

%-----------------------------------------------
\subsection{Massless twisted duality}
%-----------------------------------------------

The twisted-duality relations for the massless 
spin-2 theory around (A)dS backgrounds are 
\begin{align}
    K^C_{a[n-2]|bc} \approx \tfrac{1}{2}\,\epsilon_{a[n-2]ij}\,
    K{}^{ij|}{}_{bc}\;.
    \label{twisted_dual}
\end{align}
As usual for twisted-duality relations, the Bianchi 
identities in a formulation of the theory are mapped 
to the field equations of the dual formulation, and 
vice versa, as we now explain in details.

First, the algebraic Bianchi identity \eqref{BI} 
for the left-hand side of the twisted-duality relation 
\eqref{twisted_dual} implies that 
the trace of $K_{ab|cd}$ vanishes on-shell, which is 
the field equation \eqref{einstein1} in the metric formulation. 
The converse is true: If one takes the trace of the relation 
\eqref{twisted_dual}, the right-hand side vanishes by virtue
of the algebraic Bianchi identity \eqref{b1}. This implies 
that the trace of the left-hand side of \eqref{twisted_dual}
vanishes, which enforces the field equation \eqref{EinsteinI}
in the dual formulation. 

Second, starting again from the twisted-duality equation 
\eqref{twisted_dual}, 
the differential Bianchi identity \eqref{BianchiII2}
on the second column of $K^C$ combined with the Bianchi 
differential identity \eqref{bianchidiff} imply 
the on-shell vanishing of $K^T\,$, that is, \eqref{ZeroKT}.
Using this result, the differential Bianchi identity 
\eqref{BianchiII1} on the first column of $K^C$ gives 
the first equation of \eqref{BianchiII} that implies in its turn, 
via \eqref{twisted_dual}, the field equation \eqref{einstein2}
in the metric formulation of the massless spin-2 theory. 
The converse is also true: acting on the twisted-duality relation
\eqref{twisted_dual} with $\nabla^a$ gives identically
zero, from the right-hand side and as a consequence of 
the differential Bianchi identity \eqref{bianchidiff} for 
the curvature in the metric formulation of linearised gravity
around (A)dS. This implies the first field equation 
\eqref{EinsteinII} for the dual graviton.
Moreover, acting on \eqref{twisted_dual} with $\nabla_{\!d}$
and antisymmetrising over the three indices $\{b,c,d\}$ gives
identically zero from the right-hand side of \eqref{twisted_dual},
as a consequence of \eqref{bianchidiff}. That implies 
the field equation \eqref{ZeroKT} (and therefore the second 
field equation \eqref{BianchiII}) by virtue of the identity 
\eqref{BianchiII2}.
Finally, the field equation \eqref{einstein2} is mapped 
to the second field equation in \eqref{EinsteinII}.  

Third, the twisted-duality relation \eqref{twisted_dual} 
exactly reproduces, 
in the limit where the cosmological constant goes to zero, 
the twisted-duality relations given by Hull in 
\cite{Hull:2001iu} for linearised gravity in flat spacetime, 
see also section 4 of \cite{Bekaert:2002dt}.

%%ienc%%%%ial B%%%%%%%%%%%%%%%%%%%%%%%%%%%%%%%%%%%%%%%
\section{Partially-massless spin-2 twisted duality}\label{sec:stueckelberg}
%%%%%%%%%%%%%%%%%%%%%%%%%%%%%%%%%%%%%%%%%%%%%%%%%%%%%%

\subsection{Standard Stueckelberg formulation}

We consider the Stueckelberg Lagrangian for a partially-massless, 
symmetric spin-2 field in which both signatures are 
allowed (making AdS manifestly non-unitary at the classical 
level):
\begin{align}
{\cal L}_{PM} &= -\tfrac{1}{2}\,\nabla_{\!a} h_{bc}\nabla^a h^{bc} 
+ \nabla_{\!a} h_{bc}\nabla^c h^{ba} 
 + \tfrac{1}{2}\,\nabla_{\!a} h \nabla^a h - \nabla_{\!a} h\nabla_b h^{ab} -\tfrac{(n-1)\sigma\lambda^2}{2} \left(2 h_{ab}h^{ab}-h^2\right) \nonumber  \\
 &\qquad +\sigma\,\nabla_{\![a}A_{b]} \nabla^{[a}A^{b]} 
 + (n-1) \lambda^2 A_a A^a 
-2\widetilde{m}\, A_a \left( \nabla^a h - \nabla_{\!b}h^{ab}\right) \nonumber \\
&\qquad +\sigma \,\widetilde{m}^2 \left( h_{ab} h^{ab}- h^2 \right) \;, 
\label{LPM}
\end{align}
where the partially massless theory really appears 
in the limit
\begin{align}
    \widetilde{m}^2~\longrightarrow ~\frac{(n-2) 
    \lambda^2}{2}\;\;. 
\label{pmlim}
\end{align}
The last two lines in the expression \eqref{LPM} 
are new terms in comparison with the Lagrangian for a strictly
massless spin-2 field in (A)dS, see \eqref{LFP}.
In the limit \eqref{pmlim}, the Lagrangian ${\cal L}_{PM}$ 
is invariant, up to total derivatives, under the gauge transformations
\begin{align}
    \delta h_{ab} &= 2 \,\nabla_{\!(a}\xi_{\,b)} + \tfrac{2\widetilde{m}}{n-2}\,g_{ab}\,\epsilon\;,
    \qquad 
    \delta A_{a} = \nabla_{a}\epsilon  + 2\,\sigma\, \widetilde{m} \;\xi_a\;.
\end{align}
The quantity 
\begin{align}
    H_{ab} = h_{ab}-\tfrac{\sigma}{\widetilde{m}}\,\nabla_{(a}A_{b)}
\end{align}
is invariant under the gauge transformations with parameter 
$\xi_a\,$, but not under the gauge transformations with 
parameter $\epsilon\,$. 
A fully gauge-invariant quantity is provided by the 
antisymmetrised curl of $H_{ab}\,$. Indeed, defining
\begin{align}
    {\cal K}_{ab|c} := 2\,\nabla_{\!c}\nabla_{\![a}A_{b]} -4\sigma\lambda^2
    g_{c[a}A_{b]}-4\sigma \widetilde{m}\,\nabla_{\![a}h_{b]c}
    \equiv -4\sigma\widetilde{m}\,\nabla_{\![a}H_{b]c}\;,
    \label{invariantbis}
\end{align}
we have that ${\cal K}_{ab|c}$ is fully gauge invariant
in the partially massless limit \eqref{pmlim}, 
hence so is $\nabla_{[a}H_{b]c}\,$.
We further define the derived quantity $Q^{ab|mn}$ as follows:
\begin{align}
Q^{ab|mn} =\;&
-\tfrac{1}{2}\,\left(
\nabla^a \nabla^{[m} H^{n]b} 
- \nabla^b \nabla^{[m} H^{n]a} 
+ \nabla^m \nabla^{[a} H^{b]n} 
- \nabla^n \nabla^{[a} H^{b]m}
\right) \nonumber \\
& \qquad + (1-\tfrac{2\widetilde{m}^2}{(n-2)\lambda^2})\,
\sigma\lambda^2 \left({g}^{a[m} H^{n]b} - {g}^{b[m} H^{n]a} 
\right)\;.\label{Q}
\end{align}
It possesses the symmetries of the components of the Riemann
tensor, like $K_{ab|cd}$ in the massless case. 
The second line of the above expression is 
identically vanishing in the limit \eqref{pmlim}, so 
that $Q^{ab|mn}$ is indeed a composite object purely 
built out of the gauge-invariant quantity $\nabla_{[a}H_{b]c}\,$.
The writing that we adopted in \eqref{Q} 
facilitates the relation between $K_{ab|cd}$ and $Q_{ab|cd}\,$.
The interest in defining \eqref{Q} rests in the fact 
that the field equations for $h_{ab}$ read
\begin{align}
    \frac{\delta {\cal L}_{PM}}{\delta h^{ab}} \equiv 
    -2 \,G_{ab}\;,\quad 
        {\rm where}\quad 
G_{ab} := (Q_{ac|b}{}^{c}-\tfrac{1}{2}\,g_{ab}\,Q^{cd|}{}_{cd})\;.
\end{align}
As a consequence, the field equations 
for $h_{ab}$ imply that the curvature $Q_{ab|cd}$ is traceless
on-shell, as it was for $K_{ab|cd}$ in the 
strictly massless case. 

The Noether identities associated with the gauge parameter $\xi_a$
give the left-hand side of the field equations for the vector 
$A_a\,$:
\begin{align}
    \frac{\delta {\cal L}_{PM}}{\delta A^{a}} &\equiv
    - \frac{2\sigma}{\widetilde{m}}\;\nabla^bG_{ab}\;.\quad
\end{align}
The non-vanishing of the covariant divergence of $G_{ab}$ is 
also related to the Bianchi identity
\begin{align}
    \nabla^{[a}Q^{bc]|}{}_{mn} \equiv 
    -\frac{\widetilde{m}}{n-2} 
    \,\delta^{[a}{}_{[m}{\cal K}^{bc]|}{}_{n]}\;,
    \label{BIIpm}
\end{align}
where the gauge-invariant quantity ${\cal K}_{ab|c}$ 
was defined above in \eqref{invariantbis} and satisfies 
the identity ${\cal K}_{[ab|c]}\equiv 0\,$.
In terms of ${\cal K}_{ab|c}\,$,  the left-hand 
side of the field equations for $A_a$ reads
\begin{align}
    \frac{\delta {\cal L}_{PM}}{\delta A^a} \equiv \sigma \,
    {\cal K}_{ab|}{}^b\;,
    \label{eomA}
\end{align}
so that the field equations for $A_a$ imply that the 
curvature ${\cal K}_{ab|c}$ is traceless on-shell.

\subsection{Dual formulation}
\label{sec:stueckelberg:PM}

We now consider the dual formulation of the 
partially-massless spin-2 theory that is described by the Lagrangian 
${\cal L}_{PM}(W,U)$ in Eq.~(39) of \cite{Boulanger:2018shp}:
\be
\label{LagPM}
\begin{split}
\cL_{PM}(W,U) & = 
-\tfrac{1}{2\lambda^2} \,\nabla_{\!d} W^{bcd|a} \nabla^e W_{abe|c} 
+ \tfrac{\sigma}{\tilde{m}}\,U_{abc} \nabla_d W^{abd|c} \\
& \qquad - \tfrac{\sigma}{2(n-2)\tilde{m}^2}\, \nabla_{\!c} 
U^{abc} \nabla^d U_{abd}  - \tfrac{\lambda^2}{2 \tilde{m}^2}\,
U^{abc} U_{abc} \; . \\
\end{split}
\ee
A Lagrangian depending only on the field $W_{abc|d}$ has first been 
obtained by solving the equations of motion given by the variation of the vielbein in a first-order formulation of the partially-massless theory. In analogy with the massless case, the additional field $U^{abc}$ has then been introduced by a Stueckelberg shift.

Starting from \eqref{LagPM} one can define the following quantities 
\begin{align}
    {\cal R}_{ab|}{}^{cd} &:= 
    2\,\nabla_{\![a|}\! \left( \nabla_{\!e}W^{cde}{}_{|b]}
    - \tfrac{\sigma \lambda^2}{\tilde{m}} \, U_{|b]}{}^{cd} \right)\,,
    \\[5pt]
    {\cal K}^U_{ab|}{}^c & := 
    2\,\nabla_{\![a}\nabla^{e}U_{b]}{}^c{}_e 
    + 2(n-2) \tilde{m} \left( \nabla_{\!e}W^{ec}{}_{[a|b]} - \tfrac{\sigma \lambda^2}{\tilde{m}} \,U_{ab}{}^c \right)\,,
\end{align}
together with the successive traces
\begin{equation}
    {\cal R}_{a|}{}^{c} = {\cal R}_{ab|}{}^{cb}\;,\qquad
    {\cal R} = {\cal R}_{a|}{}^{a}\equiv 0\;,\qquad
    {\cal K}^U_{a} = {\cal K}^U_{ab|}{}^b \equiv 0\;.
\end{equation}
In a similar manner to the massless case, we introduce the traceless tensor ${\cal V}_{ab|}{}^{cd}$ according to
\begin{equation}
    {\cal V}_{ab|}{}^{cd} = {\cal R}_{ab|}{}^{cd} - 
    \tfrac{4}{n-2} \,
    \delta^{[c}{}_{[a} {\cal R}_{b]|}{}^{d]} \;,
\end{equation}
and we find that the tensors ${\cal V}_{ab|}{}^{cd}$ 
and ${\cal K}^U_{abc}$ are 
invariant under the following gauge transformations:
\begin{align}
\delta W^{bcd|}{}_{a} &= \nabla_{\!e} {\upsilon}^{bcde|}{}_{a}
+ \nabla_{\!a} \chi^{bcd} - \tfrac{3}{n-2}\, \delta_a{}^{[b} \nabla_{\!e} \chi^{cd]e} 
-\tfrac{\sigma\lambda^2}{\tilde{m}}\, \rho^{bcd}{}_{a} \;,
\\[10pt]
\delta U^{abc} &= \nabla_{\!d} \rho^{abcd} - (n-3)\tilde{m} \, \chi^{abc} \;.
\end{align}

Also in this case, we then express $W$ and $U$ in terms of their Hodge duals  
\begin{align}
 W^{abc|}{}_d &= -\tfrac{1}{(n-3)!}\,\epsilon^{e[n-3]abc}
 C_{e[n-3]|d}\;,
 & U^{abc} = -\tfrac{1}{(n-3)!}\,\epsilon^{d[n-3]abc}
 A_{d[n-3]}\;.
\end{align}
The curvature tensor for $C$ is defined, as in the massless case, 
by 
\begin{align}
{\cal K}^C_{a[n-2]|bc} = \tfrac{1}{2!}\,
\epsilon_{a[n-2]de}\,{\cal V}_{bc|}{}^{de}\;.
\end{align}
We also define the curvature $\widetilde{{\cal K}}^{a[n-2]|}{}_b$ via
\begin{align}
    {\cal K}^U{}_{ab|}{}^c = 
    (-1)^{n-1}\frac{2}{(n-2)!}\,
    \epsilon_{d[n-2][a}{}^c\,
    \widetilde{{\cal K}}^{d[n-2]|}{}_{b]}\;.
\end{align}
In order to invert this relation, we first compute
\begin{align}
    \tfrac{(-1)^n}{2}\,\epsilon^{d[n-2]ab}\,
    {\cal K}^U{}_{ab|c} &=
    \widetilde{{\cal K}}^{d[n-2]|}{}_c - (n-2)\delta^d{}_c\,
    \widetilde{{\cal K}}^{ed[n-3]|}{}_e\;
    \label{interm}
\end{align}
and take the trace of the above relation, which produces 
\begin{align}
    \widetilde{{\cal K}}^{ab[n-3]|}{}_a = \tfrac{1}{4}\,
    \epsilon^{b[n-3]cde}\,{\cal K}^U{}_{cd|e}\;.
\end{align}
Inserting this relation back in \eqref{interm} gives 
\begin{align}
    \widetilde{{\cal K}}^{a[n-2]|}{}_b = \tfrac{(-1)^n}{2}\,
    \epsilon^{a[n-3]cde}\,\Big(\delta^a{}_e\, 
    {\cal K}^U{}_{cd|b}
    - \tfrac{n-2}{2}\,\delta^{a}{}_b\,{\cal K}^U{}_{cd|e}\Big)\;.
\end{align}
Explicitly, we have
\begin{align}
    \widetilde{{\cal K}}^{a[n-2]|b} = (n-2)\,\Big(
    \nabla^b \nabla^a A^{a[n-3]} 
    + (n-2)\widetilde{m}\,\nabla^{a}C^{a[n-3]|b}
    -\sigma(n-2)\lambda^2\,g^{ab}\, A^{a[n-3]}\Big)\;,
\end{align}
which is gauge invariant under \cite{Boulanger:2018shp}
\begin{align}
\delta C_{a[n-3]|b} & = (-1)^{n-1}(n-3) \left( \nabla_{\!a} \tilde{\upsilon}_{a[n-4]|b} - \tfrac{\sigma\lambda^2}{\widetilde{m}}\, g_{ba}\, \tilde{\rho}_{a[n-4]} \right) \nonumber \\
&\qquad + \tfrac{n-3}{n-2} \left( \nabla_{\!b} \tilde{\chi}_{a[n-3]} + (-1)^n \nabla_{\!a} \tilde{\chi}_{a[n-4]b} \right)\; , \\[10pt]
\delta A_{a[n-3]} & = (n-3) \left( (-1)^{n-1}\nabla_{\!a} \tilde{\rho}_{a[n-4]} 
- \widetilde{m}\, \tilde{\chi}_{a[n-3]} \right) \;.
\end{align}
The curvatures obey the following algebraic 
Bianchi identities
\begin{align}
    {\cal K}^C{}_{a[n-2]|ab} & \equiv 0\;,
    &\widetilde{{\cal K}}_{a[n-2]|a} \equiv 0\;,
    \label{BIpm}
\end{align}
which means that ${\cal K}^C{}_{a[n-2]|bc}$ and $\widetilde{{\cal K}}_{a[n-2]|b}$ are projected on the following $GL(n)$-irreducible Young tableaux
\begin{equation*}
\ytableausetup
 {mathmode, boxsize=2}
\ytableausetup
 {mathmode, boxsize=2em}
\begin{ytableau}
 \scriptstyle a_1 & \scriptstyle b \\
 \scriptstyle a_2 & \scriptstyle c \\
\vdots\\
     \scriptstyle a_{n-2}
  \end{ytableau}
  ~~~~~\text{and}~~~~~
  \ytableausetup
 {mathmode, boxsize=2}
\ytableausetup
 {mathmode, boxsize=2em}
\begin{ytableau}
 \scriptstyle a_1 & \scriptstyle b \\
 \scriptstyle a_2\\
\vdots\\
     \scriptstyle a_{n-2}
  \end{ytableau}\quad .
  \end{equation*}
The left-hand sides of the equations of motion 
derived from the Lagrangian \eqref{LagPM} are given by
\begin{align}
    \frac{\delta \mathcal{L}_{PM}}{\delta W_{abc|}{}^d} &= \tfrac{1}{2\lambda^2} \,{\cal V}^{[ab|c]}{}_d \;,
    &\frac{\delta \mathcal{L}_{PM}}{\delta U_{abc}} = \tfrac{\sigma}{2(n-2)\tilde{m}^2} \,{\cal K}_U^{[ab|c]}\;.
\end{align}
Combining with what we obtained above, 
the field equations therefore imply 
\begin{align}
    \widetilde{{\cal K}}_{a[n-3]b|}{}^b& \approx 0\;, 
    &{\cal K}^C{}_{a[n-3]b|}{}^{bc}\approx 0\;.\label{EIpm}
\end{align}
The Bianchi identities read
\begin{align}
    \nabla_d {\cal V}_{ab|}{}^{cd} &\equiv -\tfrac{\sigma\lambda^2(n-3)}{(n-2)\widetilde{m}}\,
       {\cal K}^U{}_{ab|}{}^c\;, 
    &\nabla_{[a}{\cal V}_{bc]|}{}^{de} \equiv 
    \tfrac{2\sigma \lambda^2}{\widetilde{m}(n-2)}\,
    \delta^{[d}{}_{[a}\,{\cal K}^U{}_{bc]|}{}^{e]}\;.
\end{align}
In terms of the curvatures ${\cal K}^C$ and 
$\widetilde{{\cal K}}\,$, they become 
\begin{align}
    \nabla_a {\cal K}^C{}_{a[n-2]|}{}^{bc} & \equiv (-1)^n\,
    \tfrac{2\sigma\lambda^2(n-3)}{\widetilde{m}(n-2)}\,
    \delta^{[b}{}_a\,\widetilde{{\cal K}}_{a[n-2]|}{}^{c]}\;,
    \label{BII1pm}\\[10pt]
    \nabla^{[a}{\cal K}^C{}_{d[n-2]|}{}^{bc]} &\equiv
    -\tfrac{2\sigma\lambda^2}{\widetilde{m}}\,
    \widetilde{{\cal K}}^{[a}{}_{d[n-3]|}{}^{b}\,\delta^{c]}{}_d\;.
    \label{BII2pm}
\end{align}
By taking a trace of the Bianchi identity and using the field 
equations, one therefore deduces that
\begin{align}
    \nabla^b {\cal K}^C{}_{a[n-2]|bc} &\approx 
    (-1)^n\tfrac{(n-3)\sigma\lambda^2}{(n-2)\widetilde{m}}\,
    \widetilde{{\cal K}}_{a[n-2]|c}\;,\label{EII1pm}\\[10pt]
    \nabla^b {\cal K}^C{}_{a[n-3]b|}{}^{cd} &\approx -\tfrac{2\sigma \lambda^2}{(n-2)\widetilde{m}}\,
    \widetilde{{\cal K}}^{[c}{}_{a[n-3]|}{}^{d]}\;.
    \label{EII2pm}
\end{align}
%

%---------------------------------------------------------
\subsection{Partially-massless twisted duality}
%----------------------------------------------------------

The twisted duality that mixes the field equations and 
Bianchi identities of the two dual theories, 
the one for ${\cal L}_{PM}(h_{ab},A_a)$ on the one hand, and the 
one for ${\cal L}_{PM}(C_{a[n-3]|b},A_{a[n-3]})$ on the other hand, 
is
\begin{align}
    {\cal K}^C_{a[n-2]|bc} \approx 
    \frac{1}{2}\,\epsilon_{a[n-2]ij}\,
    Q{}^{ij|}{}_{bc}\;.
    \label{twisted_dual_pm1}
\end{align}
This equation plays the same role as \eqref{twisted_dual} 
in the strictly massless case.

What is new in the partially massless case compared to the 
massless case is that the flat limit of \eqref{twisted_dual_pm1} is not enough 
to describe all the degrees of freedom of a partially massless 
field. 
In fact, the 
twisted-duality relation \eqref{twisted_dual_pm1} also 
induces a duality relation between the curvatures 
$\widetilde{{\cal K}}_{a[n-2]|b}$ 
and ${\cal K}_{ab|c}\,$. This can be viewed by acting on 
\eqref{twisted_dual_pm1} with $\nabla_{\!a}$ and contracting the result
with $\epsilon^{a[n-1]d}\,$. One then uses \eqref{BII1pm} and 
the trace of \eqref{BIIpm}, taking into account that, on shell,
the traces of the four curvatures 
$\widetilde{{\cal K}}_{a[n-2]|b}\,$, 
${\cal K}^C_{a[n-2]|bc}\,$, $Q_{ab|cd}\,$ and 
${\cal K}_{ab|c}$ vanish. 
We obtain 
\begin{align}
    \widetilde{{\cal K}}_{a[n-2]|b} \approx \,(-1)^{n-1}\,
    \tfrac{\sigma\,\widetilde{m}^2}{4\lambda^2}\,
    \epsilon_{a[n-2]cd}\,{\cal K}^{cd|}{}_b\;,
    \label{twisted_dual_pm2}
\end{align}
where we stress that 
\eqref{twisted_dual_pm1} and \eqref{twisted_dual_pm2}
are equivalent for \emph{non-zero} cosmological constant. 

Now, taking the flat limit of \emph{both} 
\eqref{twisted_dual_pm1} and 
\eqref{twisted_dual_pm2}, we obtain two decoupled 
twisted-duality relations
for the two decoupled pairs of fields $(C_{a[n-3]|b},h_{ab})$
and  $(A_{a[n-3]},A_{a})\,$. Both together, they 
propagate the correct degrees of freedom for a partially
massless spin-2 field in the flat limit, as was found and 
discussed in section 4.3 of \cite{Boulanger:2018shp}.
The flat limit of \eqref{twisted_dual_pm2} gives 
\begin{align}
    \partial_b\widetilde{F}_{a[n-2]} \approx (-1)^n\,
    \tfrac{(n-2) \sigma}{8}\,
    \epsilon_{a[n-2]cd}\,\partial_bF^{cd}\;,
    \label{twisted_dual_pm2_flat}
\end{align}
where  $\widetilde{F}_{a[n-2]} = (n-2) \,\partial_a A_{a[n-3]}$ and 
$F_{ab}=2\,\partial_{[a}A_{b]}$ are the field strengths for 
$A_{b[n-3]}$ and $A_b\,$, respectively. In the flat limit, these 
latter quantities are gauge invariant, therefore the gradient $\partial_b$ 
on both sides of the above relation \eqref{twisted_dual_pm2_flat} can 
be stripped off to give, up to an unessential coefficient
that can be absorbed into a redefinition of $A_{a[n-3]}\,$,
the usual electric-magnetic duality between a 1-form and its 
dual $(n-3)$-form in dimension $n\,$.

As a consistency check for the second duality relation 
\eqref{twisted_dual_pm2}, one can start from the 
twisted-duality relation \eqref{twisted_dual_pm1} 
and this time take the curl of ${\cal K}^C$ on its 
second column of indices, which yields 
\begin{align}
    \nabla^{[b}{\cal K}^C{}_{a[n-2]|}{}^{cd]} \approx 
    \tfrac{1}{2}\,\epsilon_{a[n-2]ij}\,
    \nabla^{[b}Q^{cd]|}{}_{ij}\;.
\end{align}
We then use the Bianchi identities 
\eqref{BII2pm} and \eqref{BIIpm} and take a trace, 
taking into account the field equations \eqref{EIpm},
which allows us to obtain the relation 
\begin{align}
    \widetilde{{\cal K}}_{a[n-3][b|c]} \approx (-1)^{n-1}\,
    \tfrac{\sigma\, 
    \widetilde{m}^2}{4\lambda^2}\,\epsilon_{a[n-3]ij[b}
    {\cal K}^{ij|}{}_{c]}\;,
\end{align}
which is fully consistent with \eqref{twisted_dual_pm2}.

Finally, we come back to twisted-duality relation \eqref{twisted_dual_pm2} 
and gauge fix to zero both $A_a$ and $A_{a[n-3]}$ since they are
Stueckelberg fields as long as $\lambda$ is different from zero. 
In these gauges for the dual formulations, 
our second twisted-duality relation \eqref{twisted_dual_pm2} becomes
\begin{align}
(n-2)\nabla_{a}C_{a[n-3]|}{}^{b}\approx 
\tfrac{(-1)^n}{2}\,\epsilon_{a[n-2]cd}\nabla^{c}h^{db}\;,
\label{OurKurt}
\end{align}
while the first twisted-duality relation 
\eqref{twisted_dual_pm1} is just its curl, as one can 
readily check. 
This duality relation makes immediate contact with the one
proposed for the specific case $n=4$ 
in Eq.~(2.3) of \cite{Hinterbichler:2014xga}. 
Relation \eqref{OurKurt} identifies the dual curvature 
$\widetilde{F}_{ab|c}$ in \cite{Hinterbichler:2014xga} 
with $4\,\nabla_{\![a}C_{b]|c}\,$, 
the curl of the dual potential $C_{b|c}=C_{c|b}\,$.
Note that, once the Stueckelberg fields $A_{a}$ and $A_{a[n-3]}$ 
have been set to zero, one cannot take a smooth flat limit any longer
in the sense that physical degrees of freedom are lost 
in the flat limit. 

The advantage of our Stueckelberg formulation for the twisted-duality 
relation is that the identification of the 
helicity degrees of freedom is manifest and does not require any specific 
system of coordinates to be seen. 
In the original Stueckelberg formulation, 
$h_{ab}$ and $A_a$ carry the helicity two and one degrees of freedom, 
and the twisted-duality relations \eqref{twisted_dual_pm1} and 
\eqref{twisted_dual_pm2} identify these degrees of freedom
with the dual fields $C_{a[n-2]|b}$ and $A_{a[n-3]}\,$, respectively, 
in a manifestly covariant way.

\section*{Acknowledgments}

We performed or checked several computations with 
the package xTras \cite{Nutma:2013zea} 
of the suite of Mathematica packages xAct.
The work of N.B.\ has been supported in part by a FNRS PDR grant 
(number T.1025.14), while the work of A.C.\ has been supported in 
part by the NCCR SwissMAP, funded by the Swiss National Science 
Foundation. 

%\begin{appendix}
%\end{appendix}

%%%%%%%%%%%%%%%%%%%%%%%%%%%%%%%%%%%%%%%%%%%%%%%%%%%%%%%%%%%%%%%%%%%%

\end{document}